\documentclass[preprint,12pt]{elsarticle}

\usepackage{amssymb}

\usepackage{lineno}
\usepackage{subcaption}
\usepackage{natbib}
\begin{document}

\begin{frontmatter}



\title{Methods for a blind analysis of isobar data collected by the STAR collaboration}




\author{
J.~Adam$^{6}$,
L.~Adamczyk$^{2}$,
J.~R.~Adams$^{39}$,
J.~K.~Adkins$^{30}$,
G.~Agakishiev$^{28}$,
M.~M.~Aggarwal$^{40}$,
Z.~Ahammed$^{59}$,
I.~Alekseev$^{3,35}$,
D.~M.~Anderson$^{53}$,
A.~Aparin$^{28}$
% ...
}

% ...
\address{$^{2}$AGH University of Science and Technology, FPACS, Cracow 30-059, Poland}
\address{$^{3}$Alikhanov Institute for Theoretical and Experimental Physics NRC "Kurchatov Institute", Moscow 117218, Russia}
% ...
\address{$^{6}$Brookhaven National Laboratory, Upton, New York 11973}
% ...
\address{$^{28}$Joint Institute for Nuclear Research, Dubna 141 980, Russia}
% ...
\address{$^{30}$University of Kentucky, Lexington, Kentucky 40506-0055}
% ...
\address{$^{35}$National Research Nuclear University MEPhI, Moscow 115409, Russia}
% ...
\address{$^{39}$Ohio State University, Columbus, Ohio 43210}
\address{$^{40}$Panjab University, Chandigarh 160014, India}
% ...
\address{$^{53}$Texas A\&M University, College Station, Texas 77843}
% ...
\address{$^{59}$Variable Energy Cyclotron Centre, Kolkata 700064, India}
% ...

\begin{abstract}

In 2018, the STAR collaboration collected data from $_{44}^{96}Ru+_{44}^{96}Ru$ and $_{40}^{96}Zr+_{40}^{96}Zr$ at $\sqrt{s_{NN}}=200$ GeV to search for the presence of the chiral magnetic effect in collisions of nuclei. The isobar collision species alternated frequently between $_{44}^{96}Ru+_{44}^{96}Ru$ and $_{40}^{96}Zr+_{40}^{96}Zr$. In order to conduct blind analyses of studies related to the chiral magnetic effect in these isobar data, STAR developed a three-step blind analysis procedure. Analysts are initially provided a ``reference sample'' of data, comprised of a mix of events from the two species, the order of which respects time-dependent changes in run conditions. After tuning analysis codes and performing time-dependent quality assurance on the reference sample, analysts are provided a species-blind sample suitable for calculating efficiencies and corrections for individual $\approx30$-minute data-taking runs. For this sample, species-specific information is disguised, but individual output files contain data from a single isobar species. Only run-by-run corrections and code alteration subsequent to these corrections are allowed at this stage. Following these modifications, the ``frozen'' code is passed over the fully un-blind data, completing the blind analysis. As a check of the feasibility of the blind analysis procedure, analysts completed a ``mock data challenge,'' analyzing data from $Au+Au$ collisions at $\sqrt{s_{NN}}=27$ GeV, collected in 2018. The $Au+Au$ data were prepared in the same manner intended for the isobar blind data. The details of the blind analysis procedure and results from the mock data challenge are presented.

\end{abstract}




\end{frontmatter}


\section{Introduction}
\label{sec:Intro}
For more than a decade, the STAR Collaboration has been searching for evidence of chiral magnetic effects (CME) \cite{Abelev:2009ac,Abelev:2009ad,Wang:2012qs}. CME \cite{Kharzeev:2004ey,Kharzeev:2007jp} refers to the induction of an electric current ($\vec{J}_e$) by the magnetic field ($\vec{B}$) in a chiral system: $\vec{J}_e \propto \mu_5\vec{B}$. A chiral system bears a nonzero $\mu_5$, which characterizes the imbalance of right-handed and left-handed fermions in the system. The discovery of CME in high-energy heavy-ion collisions would confirm the simultaneous existence of ultra-strong magnetic fields, chiral symmetry restoration, and topological charge changing transitions in these collisions. On average, $\vec{B}$ is perpendicular to the reaction plane (${\rm \Psi_{RP}}$) that contains the impact parameter and the beam momenta. CME, therefore, will manifest a charge transport across the reaction plane.

A set of observables common to CME searches are the charge-separation fluctuations perpendicular to ${\rm \Psi_{RP}}$, e.g.~with a three-point correlator \cite{Voloshin:2004vk}, $\gamma \equiv \langle \cos(\phi_\alpha + \phi_\beta -2{\rm \Psi_{RP}}) \rangle$, where averaging is done over all particles in an event and over all events. To draw firm conclusions on the presence of CME, an effective way is needed to disentangle the signal and background contributions, the latter of which are intertwined with collective flow. Collisions of isobaric nuclei, e.g.~$_{44}^{96}Ru+_{44}^{96}Ru$ and $_{40}^{96}Zr+_{40}^{96}Zr$, present an opportunity to vary the initial magnetic field while keeping background conditions approximately the same \cite{CME}. Ruthenium-96 and Zirconium-96 each have 96 nucleons but with different numbers of protons, 44 and 40 for $Ru$ and $Zr$, respectively. Monte Carlo Glauber simulations indicate $_{44}^{96}Ru+_{44}^{96}Ru$ and $_{40}^{96}Zr+_{40}^{96}Zr$ collisions at the same beam energy are almost identical in terms of particle production \cite{Deng:2016knn}. The ratio of the multiplicity distributions from the two collision systems is consistent with unity almost everywhere, except in $0-5\%$ most central collisions, where the slightly larger charge radius of $Ru$ ($R_0 = 5.085$ fm) plays a role against that of $Zr$ ($R_0 = 5.02$ fm). CME analyses can focus on the centrality range of $20-60\%$, where the background difference due to the multiplicity are negligible. A theoretical calculation using the HIJING model \cite{Deng:2016knn} indicates the relative difference in the square of the initial magnetic field between $_{44}^{96}Ru+_{44}^{96}Ru$ and $_{40}^{96}Zr+_{40}^{96}Zr$ collisions approaches $15-18\%$ for peripheral events and $\approx13\%$ for central events. These estimates translate into a relative difference in the CME signal observable between the two isobars of $3\%$, assuming an $80\%$ background from elliptic flow, requiring a minimum of $1.2\times10^9$ events to pass the various analysis selection criteria to achieve a result of $5\sigma$ significance. Due to the small difference in the CME signal observables, of critical importance to the analysis is control of systematic uncertainties, in particular those related to detector acceptance and efficiency, which may vary in a time-dependent way.

In 2018, STAR collected data from isobar collisions, $_{44}^{96}Ru+_{44}^{96}Ru$ and $_{40}^{96}Zr+_{40}^{96}Zr$, at $\sqrt{s_{NN}}=200$ GeV. For the first time, the STAR Collaboration has implemented blind analyses of these data in studies related to CME. While blind analyses are not uncommon in particle physics, e.g.~Ref.~\cite{blindAnalysisReview}, the typical methods were not found to be suitable for the specific needs of STAR CME analyses. What follows is the description of the blind analysis procedure for the 2018 isobar collision data. The procedure was accepted by the STAR Collaboration prior to CME data-taking. While primarily relevant for CME-related studies, the opportunity for a blind analysis was open to all STAR analyses of 2018 isobar data. In identifying as a ``STAR blind analysis'' for the 2018 isobar running, analysts adhere to the following procedure. Subsequent STAR publications clearly identify as a blind analysis or an ``un-blind'' analysis according to the accepted procedure. The following procedure takes advantage of frequent switching of the isobar collision species during 2018 RHIC running to interleave isobar data samples from each species in a way that respects the time-variation of data running conditions. STAR collected 6.3 billion isobar events, evenly split between the two species, during the two months of isobar running. The RHIC isobar stores or ``fills'' typically lasted 20 hours, with STAR collecting data during 30-minute ``runs'' of the data acquisition system. Accelerator operators adjusted the beam optics throughout the 20 hour fills to maintain nearly constant collision rates, with the same target rate for the two isobar species.

\section{Blinding Techniques}
\label{sec:BlindTech}

\subsection{General principle}
\label{subsec:GenPrinc}

Blind analyses often rely on a ``reference sample'' and an inability to differentiate two or more samples or a particular sample from the reference (see Ref.~\cite{blindAnalysisReview} for a brief overview of blind analyses in particle physics). The reference sample is often used either to tune an analysis without pre-determined bias or to provide a reference for evaluating the significance of a result, e.g.~eliminating placebo effects or genetic conditions that may bias the result of medical studies.

\subsection{Considerations}
\label{subsec:Considerations}

While many possibilities exist, the blinding method for a particular analysis should be well-matched to the specific needs of that analysis. Among the specific considerations for analysis of the 2018 STAR isobar data are the following:
\begin{itemize}
\item The un-blind data should not be accessible by physics analysts prior to analysis tuning.
\item Accounting for time-dependent detector fluctuations is a critical component of analysis quality assurance (Q/A).
\item Accounting for run-by-run anomalies is a critical component of final analysis Q/A.
\item Randomizing variables within an event may severely compromise the quality of analysis Q/A and associated corrections. For example, randomizing the sign of charged particle tracks would prevent charge-dependent efficiency corrections; and randomizing particle azimuthal angle would destroy correlations from secondary decays. Because of these considerations, such methods are not retained as part of this procedure.
\item To ensure the isobar species have statistically comparable behaviors in terms of luminosity, event trigger composition, energy, vertex distribution, occupancy of tracks, etc., the 2018 RHIC run involved frequent switching of the isobar collision species.
\item With this consideration in mind, it is feasible to interleave or ``mix'' events from the two collision species in a given output data file as an efficient method to disguise the collision species.
\item Certain STAR experts, recused from blind physics analyses, may require isobar information during RHIC running to ensure data of sufficient quality to achieve target physics goals.
\item Calibration experts, who are recused from conducting blind physics analyses, may need access to un-blind data to ensure sufficiently robust calibrations and corrections to achieve the desired physics goals.
\item Runs of quality suitable for inclusion in physics analyses, e.g.~not exhibiting large detector inefficiencies, must be selected prior to the mixing of events from different species.
\end{itemize}

For the blind analysis of isobar data collected in 2018, STAR adopted a three-step blinding procedure. For the first step, analysts are provided output data files that mix events from the two isobar collision species, while respecting the time-dependence of run conditions. Analysts use this data sample to perform time-dependent Q/A of the data and to tune analysis codes. At the conclusion of these studies, analysts commit their code to a repository. In the second step, analysts are provided an ``unmixed-blind,'' sample suitable for calculating corrections that vary according to individual $\approx30$-minute data-taking runs. The run identification number are disguised, but the output data files do not mix events from different runs. Only these ``run-by-run'' corrections (e.g.~for changing detector efficiencies) and code alterations subsequent to these corrections are allowed during this step. At the conclusion of these studies, the final codes are committed to the repository, so that differences may be evaluated. After the analysis codes are verified, the final data analysis pass is completed using these final codes and the fully un-blind data released.

\subsection{Initial procedure}
\label{subsec:InitialProcedure}

Initial implementation of the analysis blinding procedure began prior to and during the 2018 RHIC run. To the extent possible, information pertaining to the isobar species was restricted during the run. Access to raw data for purposes of Q/A during the run was restricted to identified experts, approximately $5\%$ of the collaboration, recused from blind physics analyses. To the extent possible, all raw data samples were limited in size below the level needed for sensitivity to a CME signal, e.g.~less than 10000 events. Un-blind experts produced species-blind performance plots to evaluate data quality for the run in-progress.

Prior to the software production of the blind data, it was necessary to set detector calibrations and determine an appropriate list of quality data-taking runs. Due to the importance of robust calibrations to the physics analyses, these calibrations were performed by the relevant experts using un-blind data. These calibration experts were recused from participation in blinded physics analyses. Additionally, a committee was designated to determine data-taking runs of sufficient quality for inclusion in physics analyses. Members of this run selection committee were also recused from participation in blinded physics analyses. Production of the blind data commenced after calibrations and the designation of good runs.

No physics analysis groups are provided with un-blinded data prior to completion of the un-blinding procedure. 

\subsection{Blind data production}
\label{subsec:production}

In the blind production of data, the following information encoded in the data stream (DST) are obfuscated: the identification numbers for the event, its particular data-taking run, and RHIC fill; the event timestamp; the event collision species; and the hit rates for the east and west STAR zero-degree calorimeters (ZDC) \cite{STARtrig} and beam-beam counters (BBC) \cite{BBC}, as well as their coincidence and background rates. All output data files are assigned a generic name and pseudo-run-number that monotonically increases with time. 
The exact start time of a data production is not known to ensure, e.g.~that a particular pseudo-run-number is not trivially related to a particular isobar species. The mixing procedure and exact algorithm to re-assign pseudo-run numbers are encrypted and only known by two experts, who are recused from performing blind physics analyses. 
The reference sample, species-separated samples, and fully unblind samples are provided in a three-step process.

\subsection{Step-1: ``The Reference''}
\label{subsec:step1}

Analysts are initially provided output files composed of events from a mix of the two isobar species. The mixing procedure is not a priori known. As much as possible, the order of events respects temporal changes in running conditions. Events showing peculiar discrepancies from the initial Q/A are excluded from the sample, and events from the two species are only combined if the detector performance, e.g.~acceptance, was similar for the two events. Events are randomly rejected at the level of $\sim10\%$, so that ``counters,'' e.g.~trigger words, cannot be used to determine the species. Analysis code and time-dependent Q/A are tuned on this reference sample, committed to the analysis code repository, and kept unchanged at this stage. Among other aspects, this step enables extraction of time-dependent spectra for Q/A, detection of time-dependent anomalies, detection of secondary decays and measurement of peak widths relevant to momentum resolution.

\subsection{Step-2: ``The run by run Q/A sample''}
\label{subsec:step2}

After analysis of the reference data, analysts are provided an ``unmixed-blind sample'' comprised of files that obscure the true run number (and, hence, the isobar species) but do not mix events across different runs. The pseudo-run-number uniquely maps to one true run number and one (unknown) isobar species. The data are provided in such a way that a mix of files from each species appear in the same directory. As, in the first step, a fraction of events from each run is rejected to ensure that simple counting of events could not decipher the species. This sample enables species-blind run-by-run Q/A. Only run-by-run corrections and code alteration directly resulting from these corrections are allowed at this stage. The number of events provided per file is tuned so that statistics are sufficient for robust corrections but insufficient for deciphering the isobar species.


\subsection{Step-3: Full Un-blinding}
\label{subsec:fullUnblinding}

Once Q/A is complete and analyses of the run-by-run Q/A data are final, full un-blinding proceeds.
At this stage, physics results are produced with the previously tuned, vetted, and fixed analysis codes. In this data production, all information is un-blinded and restored to the data files.

\section{Implementation and Timeline for Blinded Analyses}
\label{sec:Implement}

No STAR physics analyses had access to species information prior to un-blinding.
The timelines for un-blinding are estimated by the blind analysts, who present regular updates to their respective physics working groups (PWG) to document progress and to inform adjustments to the timeline. Decisions to un-blind are based upon a review of thoroughly documented analysis procedures, codes, and analysis reports--including estimates of measurement uncertainty--by the relevant PWG.
In addition, for blind analyses of the isobar data, so-called ``godparent committees'' or ``GPCs,'' are set early and follow analyses closely throughout their development. The GPCs serve an important role in verifying that analyses are ready to proceed to the next stages of the blinding procedure.
After the step-1 data are available, blind-data analysts estimate a timeline for completing the necessary analyses for advancing to step-2. Based on this input from the analysts, management approves a date for the beginning of the second step. Analysts present regular updates to document progress. Regardless of progress, un-blinding occurs no earlier than the original estimate unless all blind analyses are deemed ready to proceed by STAR Management. Based upon the progress reports, un-blinding may be delayed to ensure the quality of the final results. An analogous timeline procedure is done for the full un-blinding.
Prior to the first un-blinding step, analysts prepare detailed notes documenting the procedures, cuts, corrections, systematic uncertainties, and criteria for any future run-by-run cuts and corrections. Prior to the second un-blinding step, analysts ensure that the documentation is updated and complete, including the run-by-run portion of analyses. Prior to each un-blinding step, analysts provide analysis codes for vetting and Q/A by the GPC in addition to the standard vetting within the physics working groups.

When the GPC is satisfied that an analysis is ready for un-blinding, analysts present the status of their analyses to the physics working group conveners and the physics analysis coordinator. As the un-blinding date approaches, analysts discuss with STAR management any need for delays to un-blinding to ensure the quality of results. If an unresolved disagreement exists between analysts, the decision to un-blind or extend the date lies with STAR management. After physics results are produced with un-blinded data, a review is conducted to verify that the frozen analysis code was used to produce the results.

While un-blinded data are not accessible to physics analyses until the blinding timeline is completed, management uses discretion in applying blinding to any calibration analysis. To ensure the integrity of calibrations, e.g.~those of the beamline and TPC \cite{tpc}, STAR calibration experts may require access to un-blind data. Without robust calibrations, the physics analyses may not be able to achieve the required precision for deciphering a CME signal. Therefore, the relevant experts are allowed access to the un-blind data for these tasks. Furthermore, access to un-blind data is restricted to these experts alone and the experts recuse themselves from participation in any blind physics analysis.

\section{Mock data challenge}
\label{sec:MDC}

\begin{figure}
    \centering
        \includegraphics[width=0.8\linewidth]{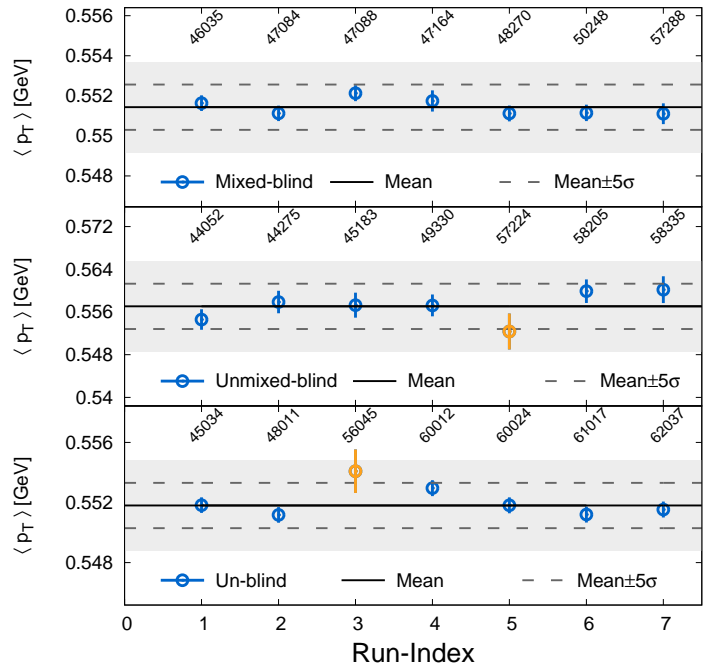}
    \caption{Mean transverse momentum of charged particle tracks associated with the primary collision vertex for three data samples: (top) ``mixed-blind'' output files containing events from a mix of data-taking runs, (middle) ``unmixed-blind'' output files containing events not mixed across data-taking runs, and (bottom) ``un-blind'' output files containing events not mixed across data-taking runs. The observable is shown as a function of an arbitrary data-taking run index with the pseudo-run-number shown in the top labels of the panel. Note that for the ``mixed-blind'' sample this index simply indicates events from the same output file.}
    \label{fig:ptVsRun}
\end{figure}

As the recommended analysis blinding procedure represents a substantial departure from that typical for STAR analyses, testing feasibility is critical. Toward this end, a ``mock data challenge'' was conducted utilizing data from $Au+Au$ collisions at $\sqrt{s_{NN}}=27$ GeV, also collected in 2018. Additionally, this exercise served as an opportunity for the software and computing team to develop, tune, and test the machinery necessary for producing the blind data samples. ``Blinded'' samples of these data were provided to analysts, utilizing the same techniques intended for blinding the isobar data. One sample was provided with output data files containing events from a mixture of data-taking runs, simulating the first stage of blinding, where data files contain a mix of isobar species. Another sample was provided using output files containing events from single data-taking runs but still blinding certain variables that in the isobar data sample could be used to identify the isobar species. For completeness, a final un-mixed sample was provided with no information obscured, simulating the fully ``un-blind'' phase of the analysis. Analysts used the two mock blind-data samples to perform quality control studies and appropriately tune analysis codes, selection cuts, and corrections. Once completed, the analysts then ran the same codes over the un-blind data sample to verify that the analysis was feasible with the given data structures and that results were appropriately consistent within the statistical differences between the samples. Example quality assurance plots for the three different samples are shown in Fig.~\ref{fig:ptVsRun}. Note that the different samples did not contain identical sets of events.

\section{After Un-blinding}
\label{sec:After}

After un-blinding, only changes to correct ``mistakes,'' defined for this purpose as errors in arithmetic or unintended departures from the approved and documented analysis procedures, are allowed.
If such a correction is made, the analysis results with the error will also be provided with a detailed explanation of the specific correction applied and why it was needed.
On a case-by-case basis, the collaboration considers announcing the result from a blind analysis simultaneously with the submission of the corresponding paper to the journal and the preprint arXiv. Regardless, only one set of ``final'' results from the blind analysis will be released, e.g.~there will be no set of ``preliminary'' results prior to the ``final'' results. All STAR publications of 2018 results state explicitly whether the analysis followed the approved STAR blinding procedure.

\section{Conclusion}
\label{sec:Conclusion}

The STAR Collaboration has developed a procedure to carry out blind analyses of isobar collision data, collected in 2018. The procedure described in this manuscript was accepted by the STAR Council in January 2018, prior to the isobar collision runs. The initial step in the procedure is an analysis of blinded data samples that interleave events from the two collision species, while the second step involves analysis of blinded data samples that do not mix events from the two collision species, followed by complete un-blinding of the data. Prior to commencing with analysis of the isobar data, a mock data challenge was successfully conducted to demonstrate the feasibility of the procedure both from an analysis standpoint and a computational standpoint. Analyses of the blind data are underway, following the procedure outlined in this manuscript.





\bibliography{ABCwriteup}





\end{document}